\documentclass[a4paper,11pt]{article}
\pdfoutput=1 

\usepackage{jcappub} 

\usepackage[T1]{fontenc} 
\usepackage{subfig}
\usepackage{siunitx}
\def\be{\begin{equation}}
\def\ee{\end{equation}}
\def\barr{\begin{array}}
\def\earr{\end{array}}
\def\bea{\begin{eqnarray}}
\def\eea{\end{eqnarray}}
\def\bfig{\begin{figure}}
\def\efig{\end{figure}}
\def\calH{{\cal H}}

\newcommand{\nn}{\nonumber}

\title{Testing Dark energy models in the light of $\sigma_8$ tension}
\author[a,b]{Gaetano Lambiase,}
\author[c]{Subhendra Mohanty,}
\author[c,d]{Ashish Narang,}
\author[c,d]{Priyank Parashari}

\affiliation[a]{Dipartimento di Fisica ``E.R. Caianiello" Universit\'a di Salerno, I-84084 Fisciano (Sa), Italy}
\affiliation[b]{INFN - Gruppo Collegato di Salerno, Italy}
\affiliation[c]{Physical Research Laboratory, Ahmedabad, 380009, India}
\affiliation[d]{Indian Institute of Technology, Gandhinagar, India}

\emailAdd{lambiase@sa.infn.it}
\emailAdd{mohanty@prl.res.in}
\emailAdd{ashish@prl.res.in}
\emailAdd{parashari@prl.res.in}

\abstract{It has been pointed out that there exists a tension in $\sigma_8-\Omega_m$ measurement between CMB and LSS observation. In this paper we show that $\sigma_8-\Omega_m$ observations can be used to test the dark energy theories.
We study two models, (1) Hu-Sawicki(HS) Model of $f(R)$ gravity and (2) Chavallier-Polarski-Linder(CPL) parametrization of dynamical dark energy (DDE), both of which satisfy the constraints from 
supernovae. We compute $\sigma_8$ consistent with the parameters of these models. We find that the well known tension in 
$\sigma_8$ between Planck CMB and large scale structure (LSS) observations is (1) exacerbated in the HS model 
and (2) somewhat alleviated in the DDE model. We illustrate the importance of the $\sigma_8$ measurements for testing 
modified gravity models. Modified gravity models change the matter power spectrum at cluster scale which also depends upon the neutrino mass. We present the bound on neutrino mass in the HS and DDE model.
}

\keywords{Dark Energy, $f(R)$ gravity, CMB, Large scale structure.}

\begin{document}
\maketitle
\flushbottom

\section{Introduction}
\label{sec:intro}
\begin{figure}[t!]
\begin{center}
 $\Lambda$CDM model \hspace{3.5cm} $\Lambda$CDM model+$\sum m_{\nu}$ \\
\hspace{-1.1cm}
\includegraphics[width=3.38in,height=2.78in,angle=0]{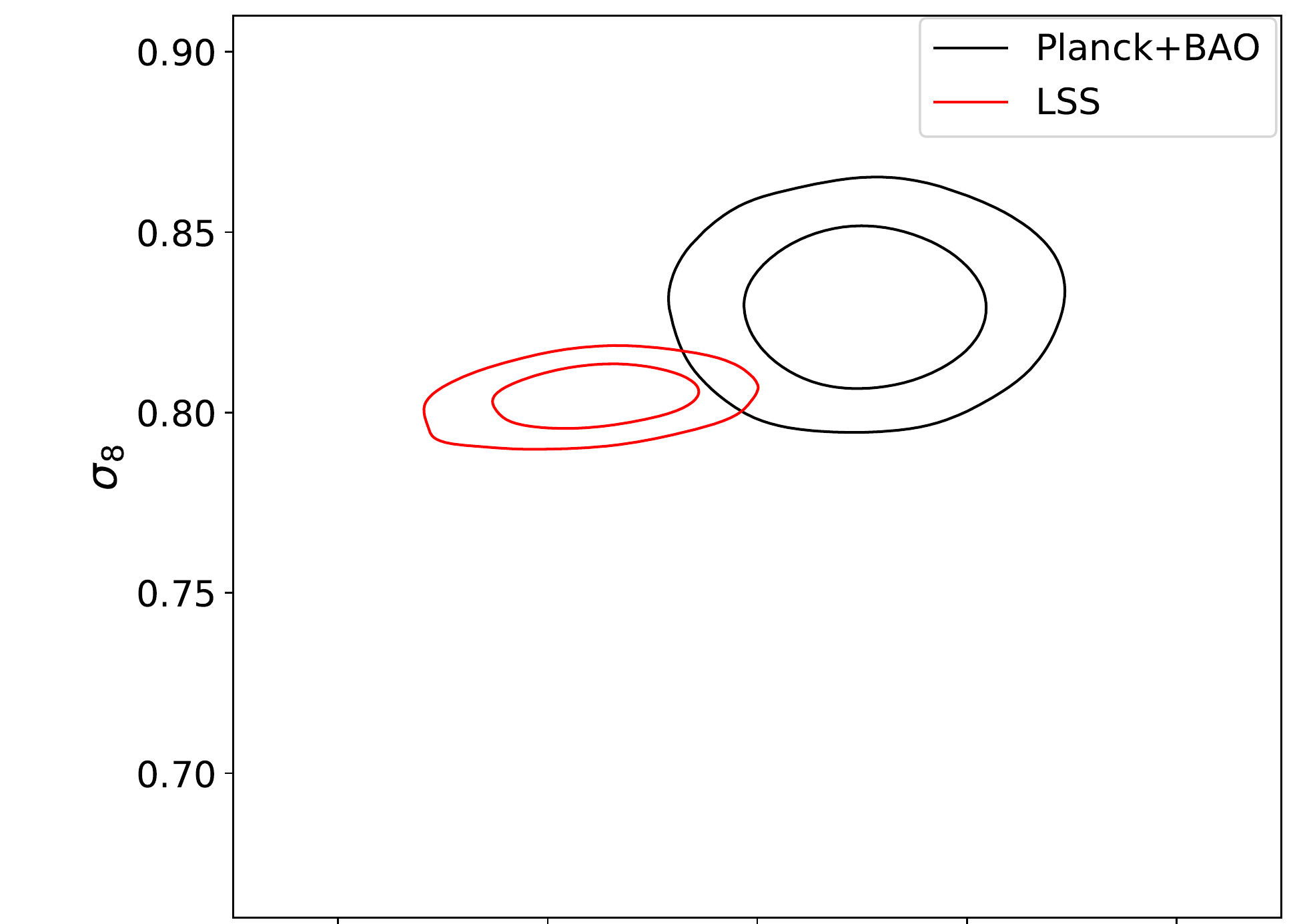}
\hspace{-0.32cm}
\includegraphics[width=2.95in,height=2.78in,angle=0]{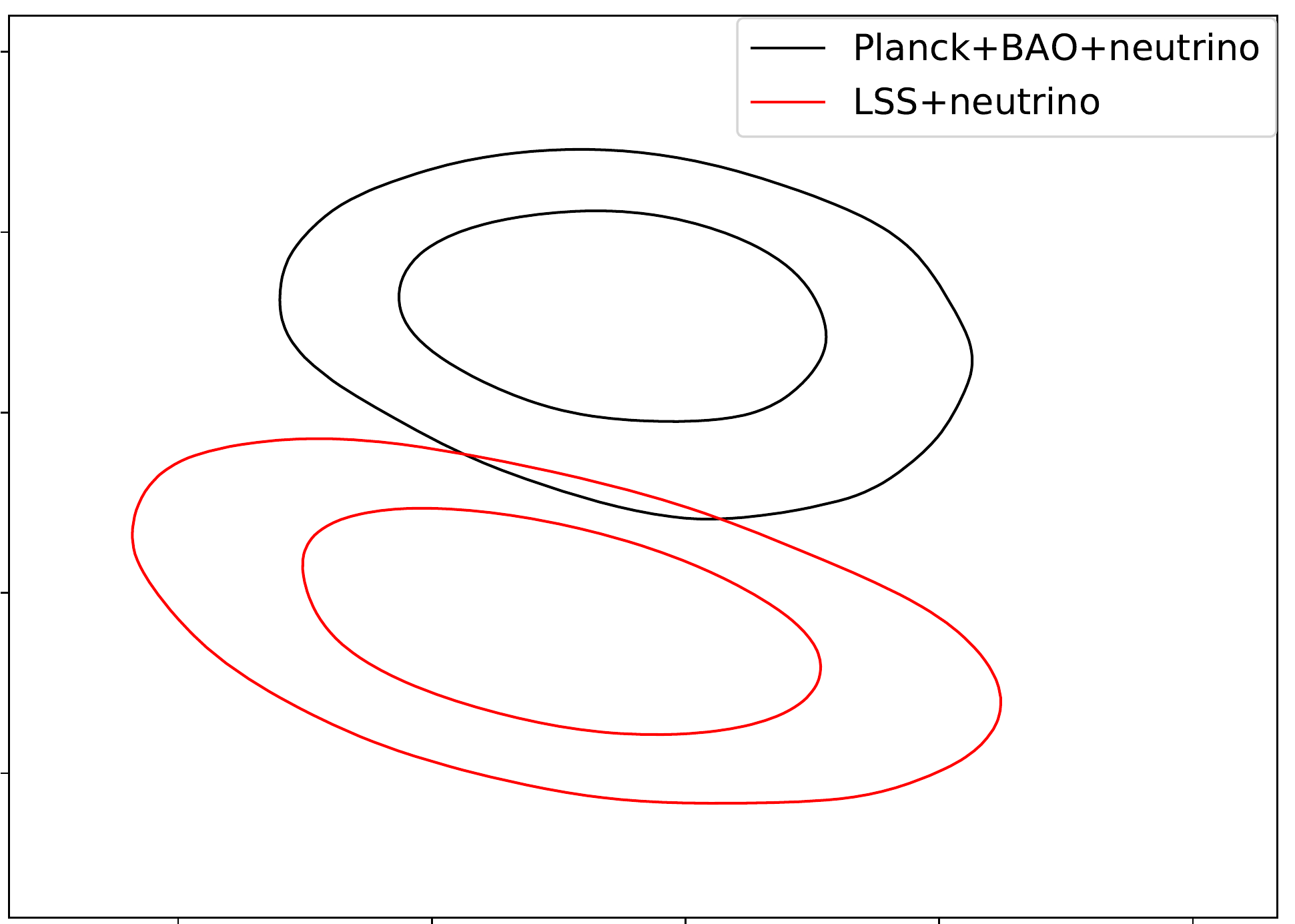}
\\
\vspace{-0.06cm}
\hspace{-1.1cm}
\includegraphics[width=3.40in,height=2.78in,angle=0]{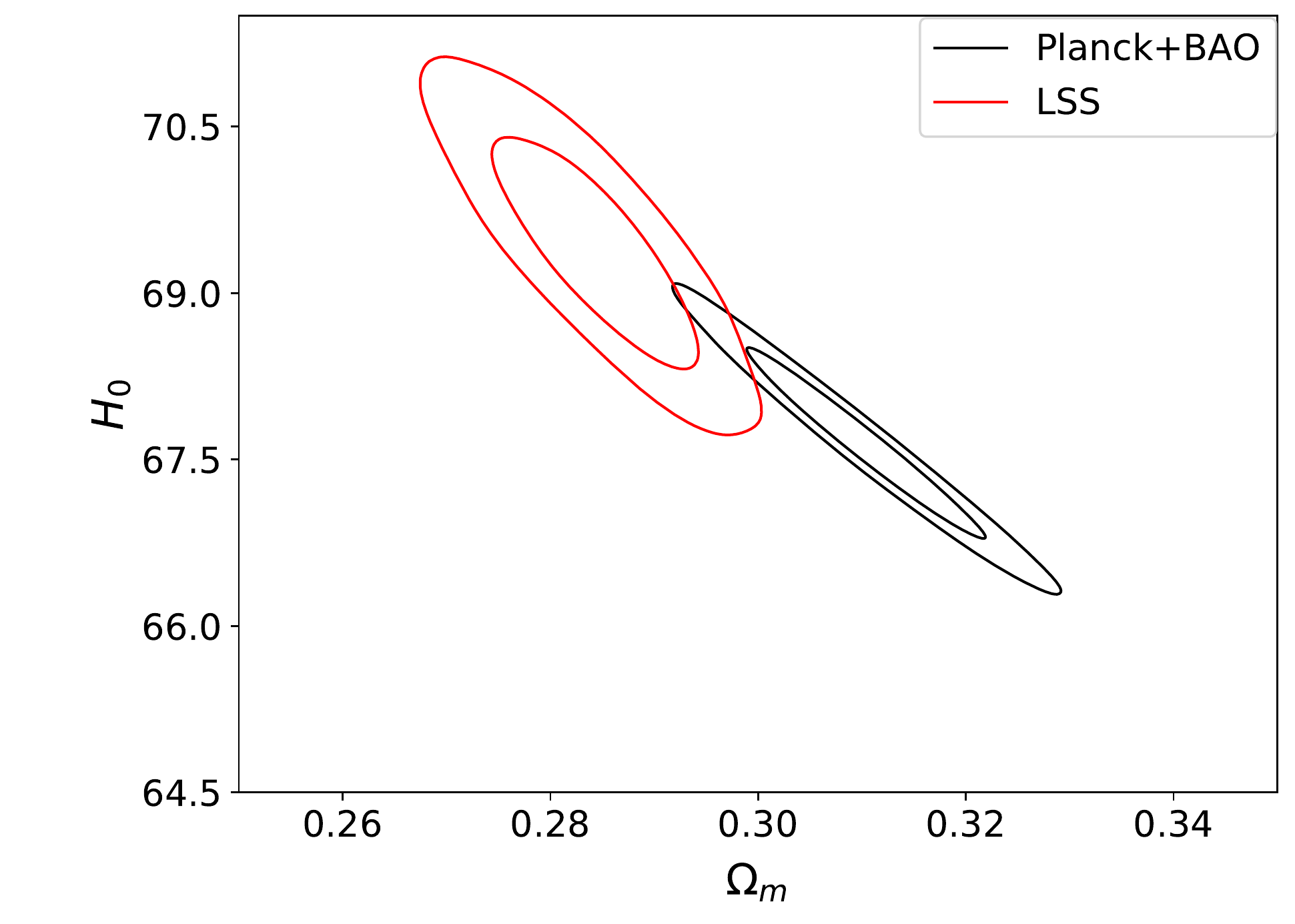}
\hspace{-0.36cm}
\includegraphics[width=2.96in,height=2.78in,angle=0]{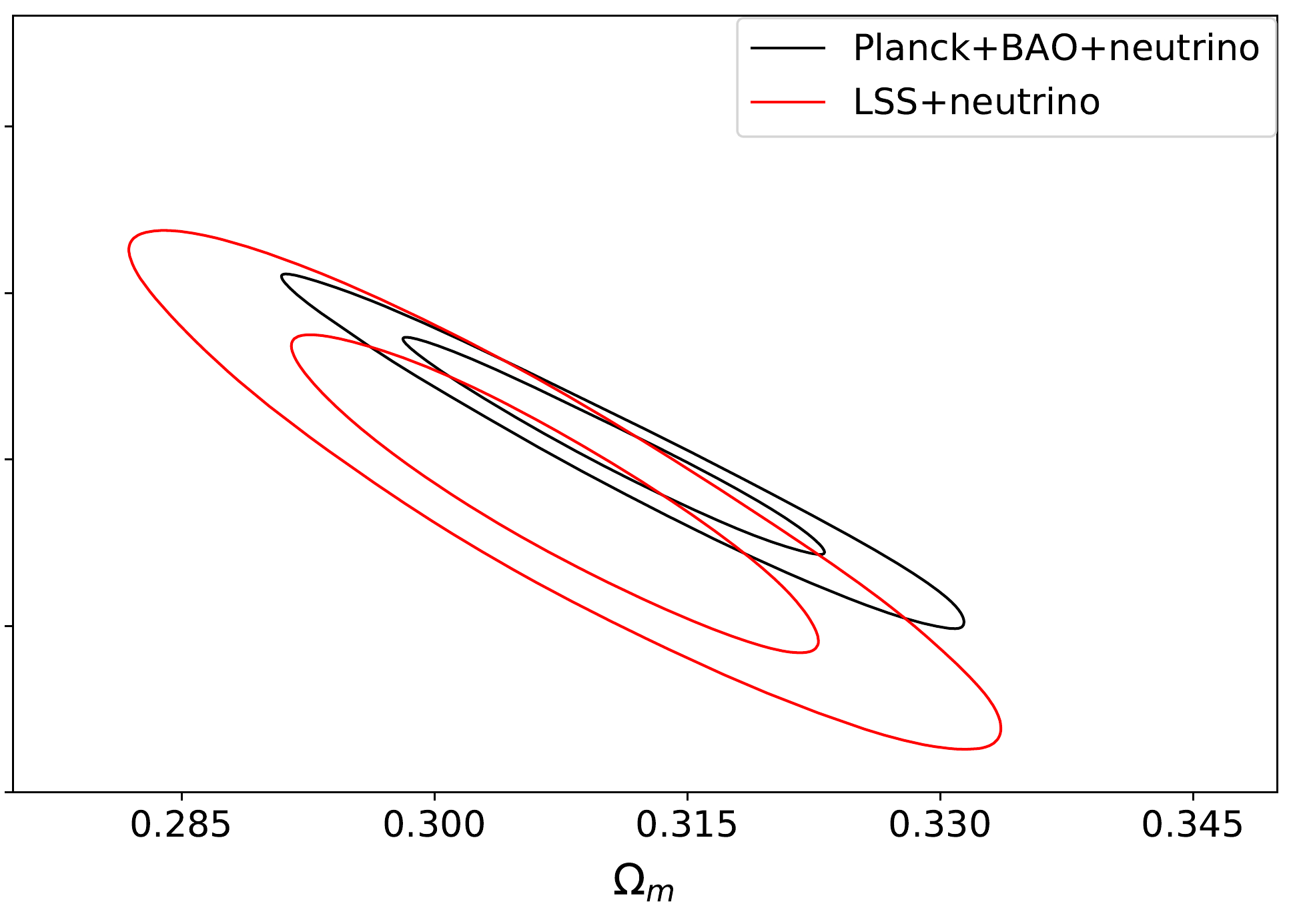} \\
(a)\hspace{6.8cm} (b) 
\caption{ (a) The discrepancy in the values of $\sigma_{8}$ and $H_{0}$ is shown here. (b) Presence of massive neutrinos resolves the $H_{0}$ tension but the $\sigma_{8}$ tension still persists.}\label{jointlambda}
\end{center}
\end{figure}

The $\Lambda$CDM model is conventional paradigm which is invoked to explain the observations of CMB temperature anisotropy and matter power 
spectrum~\cite{Ade:2015xua}. However it has been pointed out~\cite{Vikhlinin:2008ym,Macaulay:2013swa,Battye:2014qga,MacCrann:2014wfa,Aylor:2017haa,Raveri:2015maa,Lin:2017ikq} that there is some discordance between CMB 
and LSS observations. Specifically, $\sigma_8$, the r.m.s. fluctuation of density perturbations at 8 $h^{-1}$Mpc scale, 
inferred from Planck-CMB data and that from LSS observations do not agree. There have been many 
generalizations of the $\Lambda$CDM model to attempt the reconciliation between the two sets of results. For example, 
it has been shown that self interaction in dark matter-dark energy sector~\cite{Pourtsidou:2016ico,Salvatelli:2014zta,Yang:2014gza,Ko:2016uft,Ko:2016fcd,Ko:2017uyb} and several other scenarios~\cite{Gomez-Valent:2017idt, Gomez-Valent:2018nib, Ooba:2018dzf, Kazantzidis:2018rnb} can reconcile the $\sigma_8$ 
tension. There is also a tension in the inference of Hubble constant $H_0$ from CMB observations and that determined from 
LSS observations~\cite{Battye:2013xqa,Battye:2014qga}. The $H_0$ discrepancy can be resolved by including massive neutrinos~\cite{Battye:2014qga,Battye:2013xqa}, see fig. \ref{jointlambda}. In addition, 
it has been shown recently that both these anomalies can be resolved simultaneously by invoking a viscous dark matter~\cite{Anand:2017wsj}
 and effective cosmological viscosity~\cite{Anand:2017ktp}. By changing the theories of structure formation the bounds on
neutrino masses are also affected.

The main conceptual problem with $\Lambda$CDM model is that there is no explanantion to why the cosmological constant is of 
the same order as the matter density in the present epoch. One popular model which addresses this is the Hu-Sawicki 
model~\cite{Hu:2007nk} which relates the cosmological constant to the curvature in an $f(R)$ gravity theory. One may also 
take a phenomenological approach of generalising the cosmological constant to a dynamical variable and determine from 
observation how it changes in time. An example of this is the DDE model which avoids the problem of phanton crossing.

In this paper we explore the aspect of structure formation in HS Model and DDE model. The Hu-Sawicki $f(R)$ gravity 
model provides a good description of dark energy and in addition satisfies the constraints from solar system 
tests~\cite{Hu:2007nk}. We compute the power spectrum in this model and constrain the parameters with Planck-CMB and LSS data.
We find that the tension in $\sigma_8$ between Planck-CMB and LSS observations worsens in the HS model compared to 
the $\Lambda$CDM model. The second model we examine is DDE, non-phantom (equation of state $w \ge -1$) model of dark energy.
We choose the values of two model parameters in this model such that the non-phantom condition is maintained and obtain $\sigma_8$ 
from Planck-CMB and LSS data sets. We find that in the DDE model the $\sigma_8$ tension is eased as compared to $\Lambda$CDM model.

Neutrino mass cuts the power at small length scales due to free streaming. The cosmology bound on neutrino mass changes in modified gravity models. We find that the
constraint on neutrino mass $\sum m_{\nu} \le 0.157$ in $\Lambda$CDM model changes to $\sum m_{\nu} \le 0.318$ in the
HS model and $\sum m_{\nu} \le 0.116$ in the DDE model from CMB observations. Whereas, $\sum m_{\nu} =0.379^{+0.138}_{-0.148}$ 
in $\Lambda$CDM model changes to $\sum m_{\nu} =0.375^{+0.131}_{-0.144} $ in the HS model and $\sum m_{\nu} =0.276^{+0.139}_{-0.155}$ in the DDE 
model from LSS observations. We also check the $H_0$ inconsistency and find that it is being resolved on inclusion of neutrino mass in both 
these models, consistent with the earlier findings~\cite{Battye:2013xqa} that neutrino mass resolves the $H_0$ conflict.

The structure of this paper is as follows. In Sect. \ref{sec:st} we briefly discuss the Hu-Sawicki $f(R)$ model and the modification in the evolution equations. 
In Sect. \ref{sec:vde} we describe the phemomenological parametrization of DDE model. We describe the role of massive 
neutrinos in cosmology and their evolution equations in Sect. \ref{sec:massive_nu}. In Sect. \ref{sec:ps} matter power spectrum 
and it's relation to $\sigma_8$ has been discussed briefly. We also explian the efffect of HS, DDE model parameters 
and massive neutrinos on the matter power spectrum in this section followed by the 
description of data sets used and analyses done in Sect. \ref{sec:data}. We conclude with discussion in Sect. \ref{sec:conc}.

\section{ f(R) Theory: Hu-Sawicki Model}
\label{sec:st}
Scalar-tensor theories are generalized Brans-Dicke \cite{Brans:1961sx} theories. The general action for scalar-tensor theories 
is
\begin{equation}
\label{stactionein}
S_{\rm st}=\int d^4x \sqrt{-\tilde{g}} \Big(\frac{\tilde{R}}{16\pi G}-\frac{1}{2} \tilde{g}^{\mu\nu}\partial_\mu \phi \partial_\nu \phi-V(\phi)\Big)+S_m(g_{\mu\nu},\psi),
\end{equation}
where $S_m(g_{\mu\nu},\psi)$ is the action for the matter fields, ${g}_{\mu\nu}$ is Jordan frame metric and $\tilde{g}_{\mu\nu}$ is Einstein frame metric which are related by the 
conformal transformation ${g}_{\mu\nu}=A^{2}(\phi)\tilde{g}_{\mu\nu}$, and $\phi$ is the scalar field which couples to Einstein 
metric as well as to matter fields $\psi$. The scalar field brings in an additional gravitational interaction between matter
fields and the net force on a test particle modifies to 
\be
{\vec F}  = -{\vec \nabla} \Psi  -  \frac{d \ln A(\phi)}{d \phi} {\vec \nabla} \phi \ ,
\ee
and the dynamics is governed by the effective potential 
\be
V_{\rm eff}(\phi) =V(\phi) +(A(\phi)-1) \rho,
\ee
where $\Psi$ is Newtonian potential and $\rho$  is density.

The fact that scalar field couples to the matter fields would result in violations of the Einstein Equivalence Principle 
\cite{lovelock1} and signatures of this coupling would appear in non-gravitational experiments based on universality of free 
fall and local Lorentz symmetry~\cite{Kostelecky:2003fs} in the matter sector. These experiments severely constrain the presence of a scalar field 
and can be satisfied if either the coupling of the scalar field with the matter field is always very small or there is some 
mechanism to hide this interaction in the dense environments. One such mechanism is called chameleon mechanism \cite{khoury}
in which $V(\phi)$ and $A(\phi)$ are chosen in such forms that $V_{eff}(\phi)$ has density dependent minimum, 
i.e., $V_{eff}(\phi)_{min}=V_{eff}(\phi(\rho))$. The required screening will be achieved if either the coupling is very small 
at the minimum of $V_{eff}(\phi)$ or the mass of the scalar field is extremely large.

If the scalar field stays at its density dependent minimum, $\phi (\rho)$, the theory can be parametrized into two functions, the mass function $m(\rho)$ and the coupling $\beta (\rho)$ at the minimum of the potential\cite{Brax:2012gr,Brax:2011aw}
\be
\frac{\phi (\rho)-\phi_c}{m_{\rm Pl}}= \frac{1}{m_{\rm Pl}^2}\int_{\rho}^{\rho_c} d\rho \frac{\beta (\rho) }{m^2(\rho)},
\ee
where $m_{\rm Pl}$ is the Planck mass and mass of the scalar field $m(\rho)$ and the coupling parameter $\beta(\rho)$  are respectively given as
\be
m^2 (\rho)= \frac{d^2 V_{\rm eff}}{d\phi^2}\vert_{\phi=\phi (\rho)}
\ee
\be
\beta (\rho)= m_{\rm Pl} \frac{d\ln A}{d\phi}\vert_{\phi=\phi(\rho)}.
\ee
Using the evolution of the matter density given by  $\rho(a)=\rho_{0}a^{-3}$, $m(\rho)$ and $\beta(\rho)$ can be represented 
as functions of scale factor $a$, i.e, $m(a)$ and $\beta(a)$.

As discussed in Sect. \ref{sec:intro} that dark energy can be explained alternatively by modified gravity. Simplest modified gravity 
model is the $f(R)$ gravity. In general relativity (GR) Lagrangian deisity is given by Ricci scalar $R$, whereas it 
is a non linear function of $R$ in the $f(R)$ gravity. Hence the action for an $f(R)$ theory is given as
\be
 S = {1\over 16\pi G} \int d^4x \sqrt{-g}\,( f(R)) + S_m(g_{\mu\nu},\psi),
 \label{eq:action}
\ee 
where $f(R) $ is a non linear function of $R$. The scalar degree of freedom in the $f(R)$ theories has been utilized as the quintessence field to explain DE.
It has been shown~\cite{Sotiriou:2006hs, DeFelice:2010aj} that $f(R)$ theory is the equivalent to a scalar-tensor theory with an equivalence relation
\be
f_R = e^{-2 \beta_0 \phi_R/m_{\rm Pl}}, \label{eq:equi-F(R)}
\ee
and potential corresponding to extra scalar degree of freedom
\be
V(\phi_R) = {m_{\rm Pl}^2\over 2} {Rf_R - f(R) \over f_R^2}\label{eq:equi-V(phi)},
\ee
where $f_R = \partial f / \partial R$. There are many form of $f(R)$ proposed which explain the type Ia supernovae 
observation. In this paper, we consider the Hu-Sawicki model, which explains DE while evading the stringent tests from solar system observations. In HS model the 
modification in the action is given as
\be 
f(R) = R -  2 \Lambda - {f_{R_{0}}\over n}{R_0^{n+1} \over R^{n}},\label{eq:model_HU}
\ee
where $R \geq R_0$ and $R_0$ is the curvature at present. Here $f_{R_{0}}$ and $n$ are the free parameters of the HS model.
Using equivalence relation \ref{eq:equi-F(R)} and eq. \ref{eq:equi-V(phi)}, we find that
 \be
 {R_0\over R} \approx \left({-2 \beta_0 \phi_R \over m_{\rm Pl} f_{R_{0}}} \right)^ {1/(n+1)} 
\ee
 and
 \be
 V(\phi_R) = \Lambda + {n+1\over n} f_{R_{0}} R_0 \left({-2 \beta_0 \phi_R \over m_{\rm Pl} f_{R_{0}}} \right)^ {n/(n+1)}
 \ee
 
 The coupling function $\beta(a)$ is constant for all the $f(R)$ models $i.e \,\,\beta(a) = {1\over\sqrt{6}}$ , whereas the mass function is a model dependent 
 quantity ~\cite{Brax:2011aw,Brax:2012gr,Brax:2013fna}. In particular for the HS model, for which form of $f(R)$ is given by eq. \ref{eq:model_HU}, 
 we have mass function
\be 
m(a) = m_0 \left({4\Omega_{\Lambda} + \Omega_m a^{-3} \over 4\Omega_{\Lambda} + \Omega_m } \right)^{(n+2)/2},
\ee
with
\be
m_0 = H_0 \sqrt{{{4\Omega_{\Lambda} + \Omega_m \over (n+1)f_{R_{0}}} }},
\ee
These parameters contains
all the imformation of the model,  where $\Omega_{\Lambda}$ and $\Omega_m$ are the matter density fraction for dark energy and matter today. In the next subsection, we will derive the evolution equations in terms of these parameters. 

\subsection{Evolution Equations}
\label{sec:evo} 
In GR the evolution of metric perturbation potentials and density perturbations is given by the
following linearized equations,
\be
k^{2} \Phi = -4 \pi G a^{2} \rho \delta
\ee
\be
k^{2}(\Phi-\Psi) = 12 \pi G a^{2} (\rho+P)\sigma,
\ee
\be 
\delta'' + \calH \delta' - 4\pi G a^2 \rho \delta = 0
\ee
Where $'$ denotes the derivative with respect to the conformal time, $\delta$ is the comoving density contrast and $\Phi$ and $\Psi$ are the space-time dependent perturbations to the FRW
metric, 
\begin{equation}
ds^2= -(1+2\Phi)dt^2  + a^2(t) (1-2\Psi) \delta_{ij} dx^idx^j.\label{eq:metric} 
\end{equation}
In the modified gravity models and other dark energy models these relation can be different. To incorporate the possible 
deviations from $\Lambda$CDM evolution there are several parametrization \cite{silvestri,hojjati,song,bertschinger,zhang} present in the literature. We use the following parametrization which was introduced in \cite{silvestri}
\be
k^{2} \Psi = -4 \pi G a^{2} \mu(k,a) \rho \delta
\ee
\be
\frac{\Phi}{\Psi}=\gamma(k,a),
\ee
where $\mu(k,a)$ and $\gamma(k,a)$ are two scale and time dependent functions introduced to incorporate any modified theory
of gravity. Note the appearance of $\Psi$ instead of $\Phi$ in the first equation. In the quasi-static approximation 
$\mu(k,a)$ and $\gamma(k,a)$ can be expressed as \cite{Brax:2012gr}
\be
\mu(k,a)=A^{2}(\phi)(1+\epsilon(k,a)),
\ee
\be
\gamma(k,a)=\frac{1-\epsilon(k,a)}{1+\epsilon(k,a)},
\ee
where 
\be
\epsilon(k,a)=\frac{2 \beta^{2}(a)}{1+m^{2}(a)\frac{a^{2}}{k^{2}}}.
\ee
Modification in the evolution of $\Psi$ and $\Phi$ in turn modifies the evolution of matter perturbation to
as
\be 
\delta'' + \calH \delta' - \frac{3}{2} \Omega_m \calH^2 \mu(k,a) \delta = 0
\ee
where $\calH = a'/a$.

\section{Dynamical Dark Energy model}
\label{sec:vde}
The current measurements of cosmic expansion, indicate that the present Universe is dominated by dark energy (DE). The most common dark energy candidate is cosmological constant $\Lambda$ representing a constant energy density occupying the space homogeneously. The equation of state parameter for DE in cosmological constant model is $w_{DE}=\frac{P_{DE}}{\rho_{DE}}=-1$. However a constant $\Lambda$ makes the near coincidence of $\Omega_\Lambda$ and $\Omega_m$ in the present epoch hard to explain naturally. This gives way for other models of DE such as quintessence\cite{PhysRevD.35.2339,PhysRevD.37.3406,Chiba:1997ej}, interacting dark energy\cite{int} and phenomenological parametrization of DE such as DDE
\cite{cp,linder,barboza,jassal,wetterich}. In the phenomenological DE models the equation of state parameter is taken to be a variable, dependent on the scale factor (equivalently redshift),i.e.,
\be
w(z)=\sum_{n} w_n x(z),
\ee
where $w_n$ are parameters fixed by observations and $x(z)$ is function of redshift. The most commonly followed $w(z)$ dependence are phantom fields($w(z)<-1$) and non phantom field($-1 \leq w(z) \leq 1$). In this paper we use the Chavallier-Polarski-Linder(CPL)\cite{cp,linder} parametrization of DDE. The equation of state parameter for DE in CPL parametrization is
\be
w_{\rm DE}(z)=w_{0}+w_{a}\frac{z}{z+1},
\label{eq:vde}
\ee
where $w_{0}$ and $w_{a}$ are the CPL parameters. Choosing $w_{0}=-1$ and $w_{a}=0$ eq. \ref{eq:vde} gives back the $\Lambda$CDM model. As a result of this parametrization the evolution of DE density fraction is given by the equation
\be
\Omega_{\rm DE}(z)=\Omega_{\rm DE,0}(1+z)^{3(1+w_{o}+w_{a})}e^{-3 w_{a}\frac{z}{z+1}},
\ee
where $\Omega_{\rm DE,0}$ is the DE density at present. 
\section{Massive neutrino in cosmology}
\label{sec:massive_nu}
 Neutrinos play an important role in the evolution of the Universe.
 Several neutrino experiments have established that neutrinos are massive. Massive neutrinos can affect the background as well as
 matter perturbation which in turn can leave its imprint on cosmological observations. In the early universe, neutrinos are 
 relativistic and interact weakly with other particles. As the temperature of the Universe decreases, the weak interaction 
 rate becomes less than the Hubble expansion rate of the Universe and neutrinos decouple from rest of the plasma. Since 
 neutrinos are relativistic, their energy density after decoupling is given~\cite{Lesgourgues:2006nd,lesgourgues2013neutrino}
 \bea
 \rho_{\nu} = \left[{7\over 8}\left(4\over 11\right)^{4/3}N_{\rm eff}\right]\rho_\gamma\, ,
 \eea
 where $\rho_\gamma$ is the photon energy density. $N_{\rm eff}$ is the effective number of 
 relativistic neutrinos at early times and its value is equal to 3.046~\cite{Mangano:2005cc}. When the temperature of the 
 Universe goes below the mass of the neutrinos, they turn into non-relativistic particles.The energy density fraction of 
 neutrinos in the present universe depends on the sum of their masses and is given as
 \bea
 \Omega_\nu = \frac{\sum m_{\nu}}{\rm eV}{1\over 93.1 h^2}\, ,
 \eea
 where $\sum m_{\nu}$ is the sum of neutrino masses. Neutrinos in the present Universe contribute a very small fraction of energy density however they can affect the formation of structure at large scales.
 
 After neutrinos decouple, they behave as collisionless fluid with individual particles streaming freely. The free 
 streaming length is equal to the Hubble radius for the relativistic neutrinos, whereas non-relativistic neutrinos stream
 freely on the scales $k > k_{\rm fs}$, where $k_{\rm fs}$ is the neutrino free-streaming scale. On the scales 
 $k > k_{\rm fs}$, the free-streaming of the neutrinos damp the neutrino density fluctuations and suppress the power in the
 matter power spectrum. On the other hand neutrinos behave like cold dark matter perturbations on the scales 
 $k < k_{\rm fs}$.~\cite{Lesgourgues:2006nd,lesgourgues2013neutrino} 

 \subsection{Evolution equations for massive neutrinos}
\label{sec:evo} 

Massive neutrinos obey the collisionless Boltzmann equation, therefore we solve the Boltzmann equation for the neutrinos to
get their evolution equations. The energy momentum tensor for neutrinos is given as
\be
T_{\mu \nu} = \int dP_1 dP_2 dP_3 (-g)^{-1/2} {P_\mu P\nu \over P^0} f(x^i,P_j,\tau)\, ,
\label{eq:em_tensor}
\ee
where $f(x^i,P_j,\tau)$ and $P_\mu$ are the distribution function and the four momentum of neutrinos respectively. We expand 
the distribution function around the zeroth-order distribution function $f_0$ as
\be
f(x^i,P_j,\tau) = f_0 (q)[1 + \chi(x^i,P_j,\tau)] ,
\label{eq:dis-func}
\ee
 where $\chi$ is the perturbation in the distribution function. Using \ref{eq:em_tensor} in \ref{eq:dis-func} and equating the zeroth order terms, we get the unperturbed energy density and pressure for 
 neutrinos
\be
\bar \rho = 4\pi\,a^{-4}\int q^2dq \epsilon f_0(q), ~~~~ 
\bar P = {4\pi a^{-4}\over 3}\int q^2 dq {q^2\over \epsilon} f_0(q).
\ee
 Similarly, We get the purterbed quantities by equating the first order
terms
\be
\delta \rho = 4\pi\,a^{-4}\int q^2dq \epsilon f_0(q) \chi, ~~~~ 
\delta P = {4\pi a^{-4}\over 3}\int q^2 dq {q^2\over \epsilon} f_0(q)\chi. 
\label{eq:pert_rho_nu}
\ee
\be
\delta T^0_{i} = 4\pi\,a^{-4}\int q^2dq qn_i f_0(q)\chi, ~~~~ 
\delta \Sigma^i_j = {4\pi a^{-4}\over 3}\int q^2 dq {q^2\over \epsilon}(n_i n_j - {1\over 3}\delta_{ij}) f_0(q)\chi,
\label{eq:pert_T_ij_nu}
\ee
where $q_i = qn_i$ is the co-moving momentum and $\epsilon = \epsilon(q,\tau) = \sqrt{q^2 + m_\nu^2 a^2}$. It is clear from eqs. \ref{eq:pert_rho_nu} and
\ref{eq:pert_T_ij_nu} that we can not simply integrate out the $q$ dependence as $\epsilon$ is the function of both 
$\tau$ and $q$. Hence, we will use the Legendre series expansion of the perturbation $\chi$ to get the perturbed evolution
equations for the massive neutrino. Legendre series expansion of the perturbation $\chi$ is given as
\be
\chi(\vec k,\hat n, q,\tau) = \sum_{l =0}^{\infty} (-i)^l (2l +1)\,\chi_l (\vec k, q,\tau) P_l(\hat k . \hat n)\, .
\label{eq:Legendre}
\ee
Using eq. \ref{eq:Legendre} in the eqs. \ref{eq:pert_rho_nu} and \ref{eq:pert_T_ij_nu}, we get the perturbed evolution
equations for the massive neutrino ~\cite{Ma:1995ey}
\bea
\delta \rho_h& =& 4 \pi a^{-4} \int q^2 dq \epsilon f_0(q) \chi_0\, ,\nn\\ 
\delta P_h& =& {4 \pi \over 3} a^{-4} \int q^2 dq {q^2 \over \epsilon} f_0(q) \chi_0\, ,\nn\\
(\bar{\rho}_h + \bar{P}_h) \theta_h & =& 4 \pi k a^{-4} \int q^2 dq q f_0(q) \chi_1\, ,\nn\\ 
(\bar{\rho}_h + \bar{P}_h) \sigma_h & =& {8 \pi \over 3} a^{-4} \int q^2 dq {q^2 \over \epsilon} f_0(q) \chi_2\, 
\label{eq:pert_nu_stress},
\eea
where the Boltzmann equation governs the evolution of $\chi_l$. In the Newtonian gauge Boltzmann equations for 
$\chi_l$ are given as

\bea\label{eq:boltz-nu}
\dot{\chi}_0 &= & -{qk\over \epsilon} \chi_1 - \dot{\Phi} {d \ln f_0 \over d \ln q}\, ,\nn\\
\dot{\chi}_1 &= & {qk\over 3 \epsilon}(\chi_0 - 2\chi_2) - {\epsilon k \over 3 q } \Psi {d \ln f_0 \over d \ln q}\, ,\nn\\
\dot{\chi}_l &= & {qk\over (2l + 1) \epsilon} [l \chi_{l-1} - (l+1)\chi_{l+1}]\, , \hspace{0.6cm} {\rm for,}\,\,  l\geq2.
\eea

\section{Matter power spectrum and $\sigma_8$}
\label{sec:ps}
In this section we discuss the effect of massive neutrinos, HS model parameters and DDE model parameters on the matter 
power specturum and $\sigma_8$. Matter power spectrum is a 
scale dependent quantity defined as the two-point correlation function of matter density, $P(k)=k^{n_s}T^2(k) D^2(a)$. Where $T(k)$ is the matter transfer 
function, $D(a)$ is the linear growth factor and $n_s$ is the tilt of the primordial power spectrum. Also, the r.m.s.
fluctuation of density perturbations in a sphere of radius $r$ is defined as

\begin{equation}
 \sigma(r,z) = \left[\frac{1}{2\pi^2} \int_0^{\infty} dk k^2 P(k,z)|W(kr)|^2\right]^{1/2},
\end{equation}
where $r$ is related to mass by $r = (3M/4 \pi \rho_m(z=0))^{1/3}$ with $\rho_m(z=0)$ being the matter density of the Universe at 
present epoch. Here $W(kr) = 3(\sin kr - kr \cos kr)/(kr)^3$ is the filter function. This is a scale dependent quantity.
The r.m.s. fluctuation of density perturbations on scale  8 $h^{-1}$Mpc is called $\sigma_8(z)$. 

\begin{figure}[h!]
\begin{center}
\includegraphics[width=4.2in,height=3.2in,angle=0]{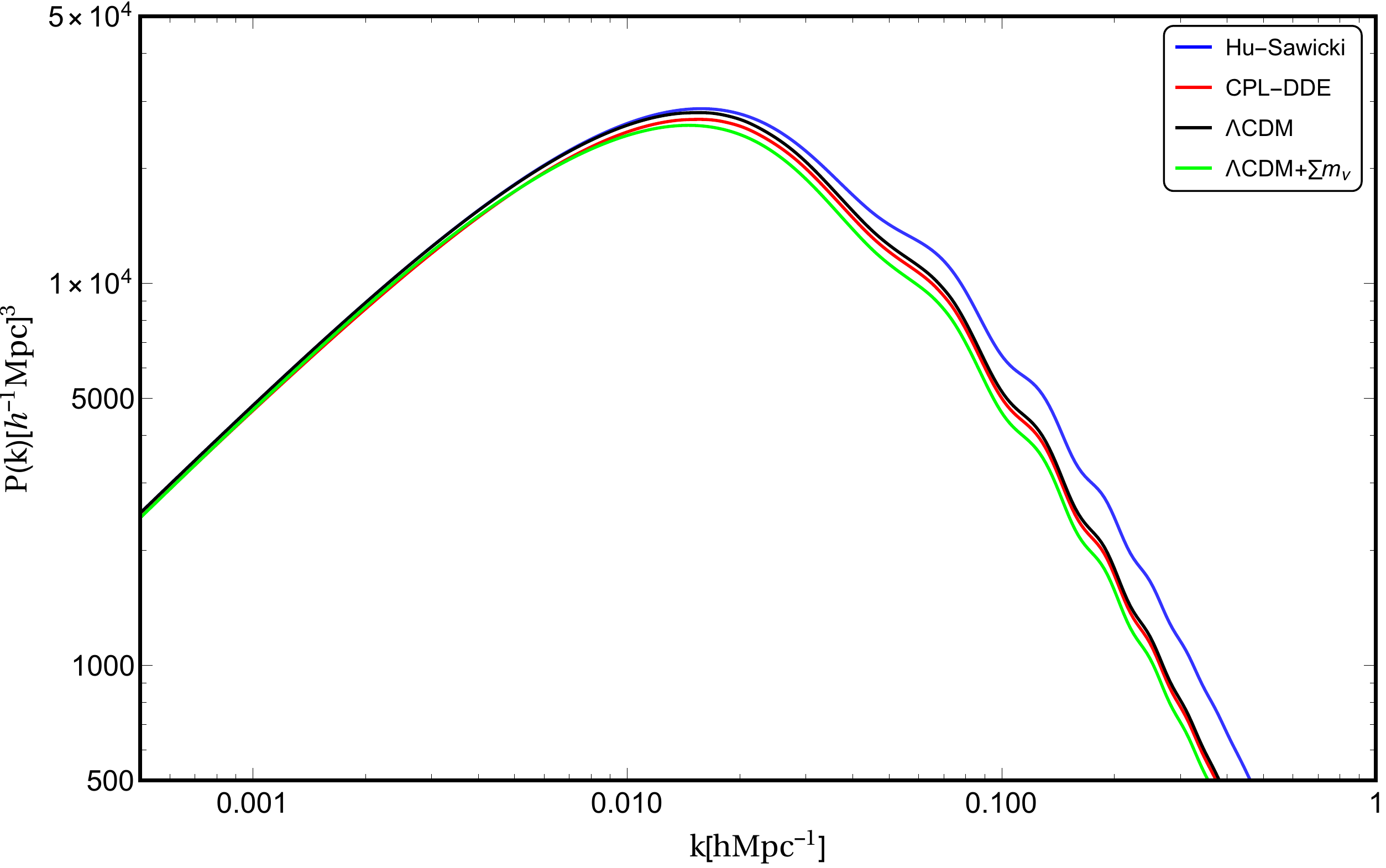}~~
\caption{Matter power spectrum in HS, DDE and $\Lambda$CDM model.}\label{fig:ps}
\end{center}
\end{figure}

We use CAMB~\cite{Lewis:1999bs} to generate 
the matter power spectrum for DDE model, whereas we use MGCAMB~\cite{hojjati,silvestri} to obtain the matter power spectrum for HS model. In 
order to see the effect of modified gravity models and massive neutrinos we plot matter power spectrum for some bench
mark values of $\sum m_{\nu}$, HS model parameters and DDE model parameters. The power spectrums are shown in 
fig. \ref{fig:ps}.
\begin{itemize}
\item As we discussed in Sect. \ref{sec:massive_nu}, 
massive neutrinos stream freely on the scales $k > k_{\rm fs}$ and they can escape out of the high density regions 
on those scales. The perurbations on length scales smaller than neutrino free streaming length will be 
washed out and therefore power spectrum gets suppress on these scales. Neutrino mass cuts the power at length scales even larger than the 8 $h^{-1}$Mpc which requires a large $\Omega_m$ which in turn disfavours the compatibility of $\sigma_8-\Omega_m$ between the two observations.
\item DDE cuts the power spectrum at all length scales. Since, in the DDE model, dark energy density increases
with the redshift, therefore, in the early time when the dark energy density is large, the power cut is more prominent at 
small scales.
\item On the other hand, the power spectrum gets 
affected in an opposite manner for HS model as the power increases slightly on small length scales.
\end{itemize}

\begin{figure}[!tbp]
\begin{center}
 HS Model \hspace{4.5cm} DDE Model \\
\hspace{-1.1cm}
\includegraphics[width=3.38in,height=2.78in,angle=0]{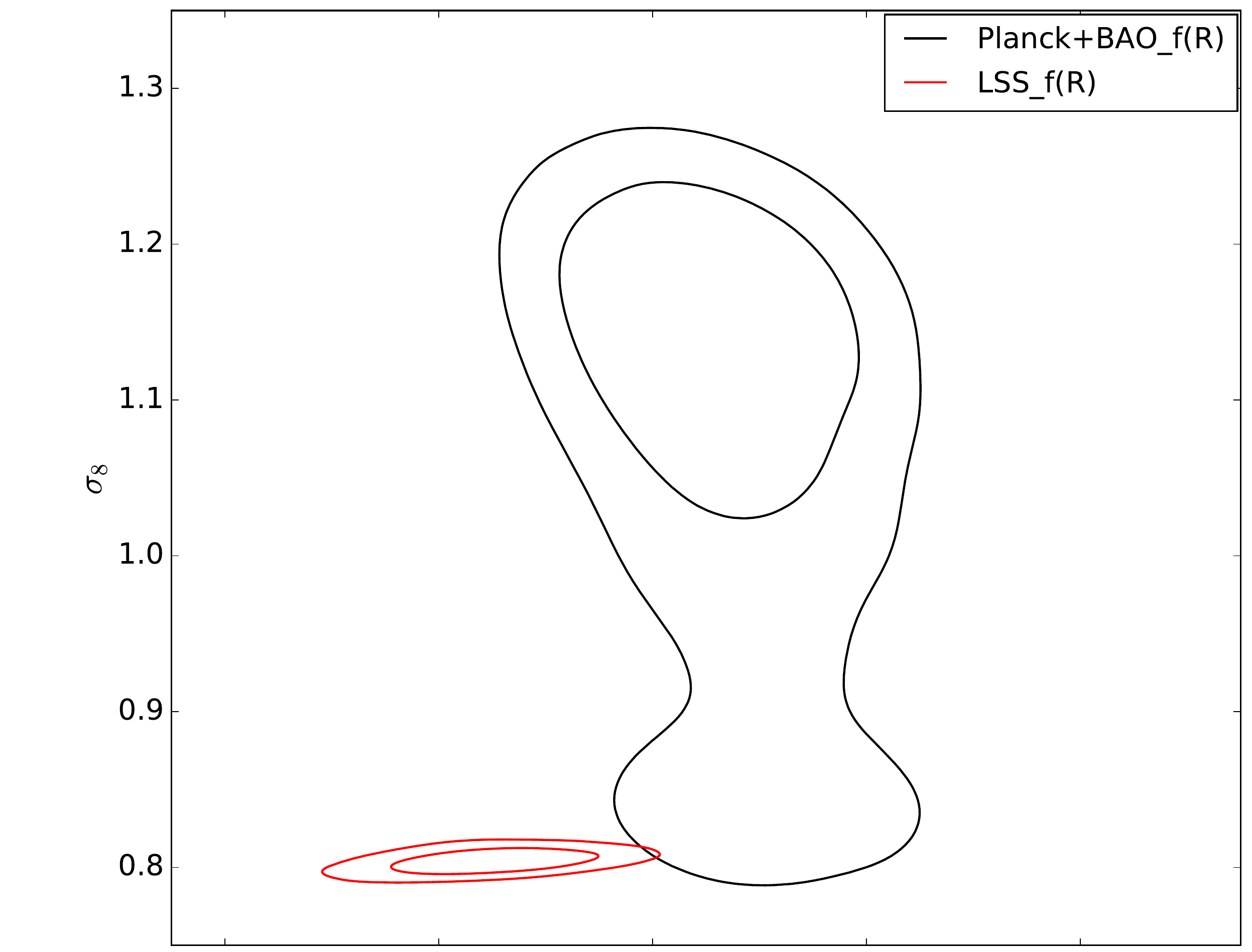}
\hspace{-0.32cm}
\includegraphics[width=2.95in,height=2.78in,angle=0]{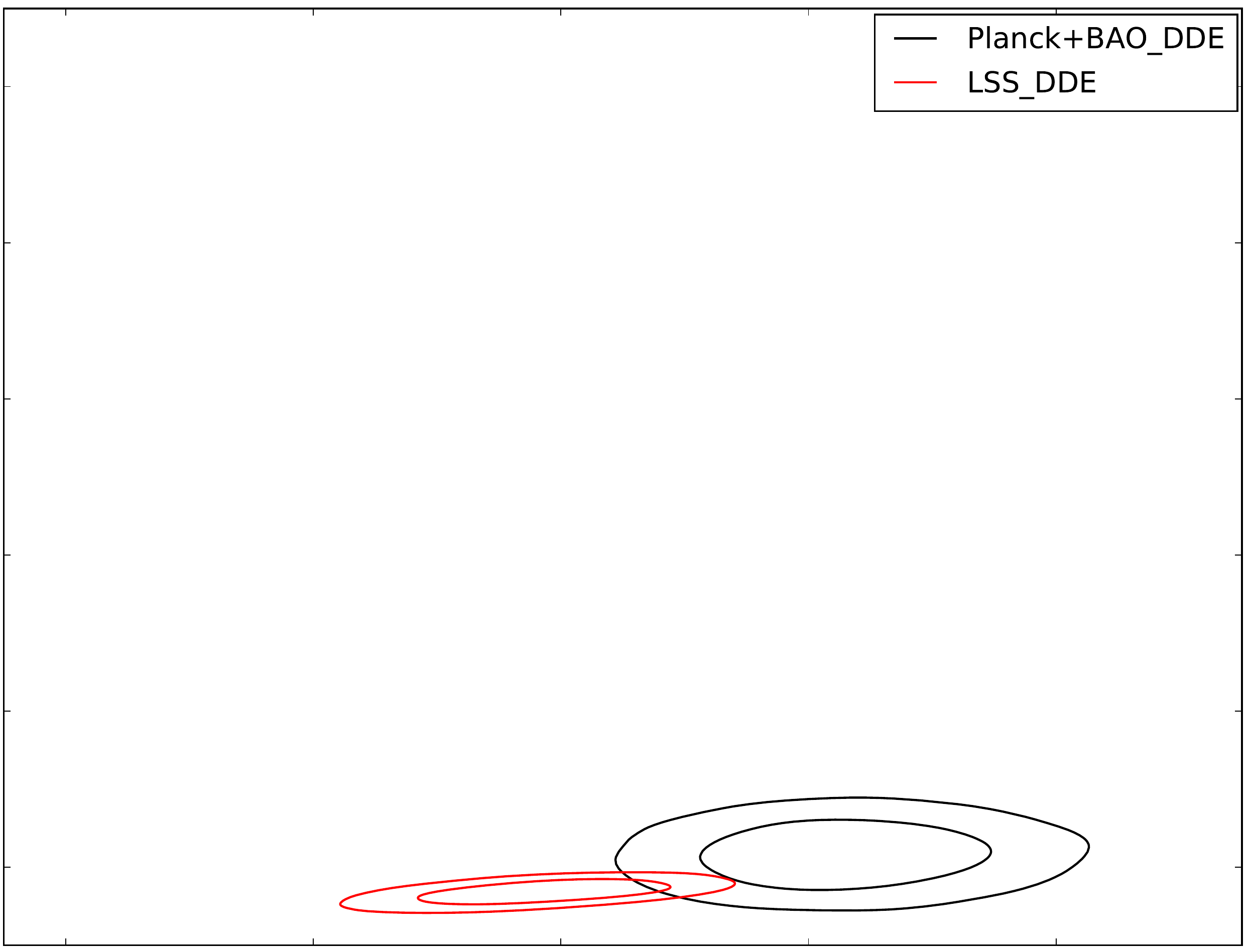}
\\
\vspace{-0.07cm}
\hspace{-1.1cm}
\includegraphics[width=3.38in,height=2.78in,angle=0]{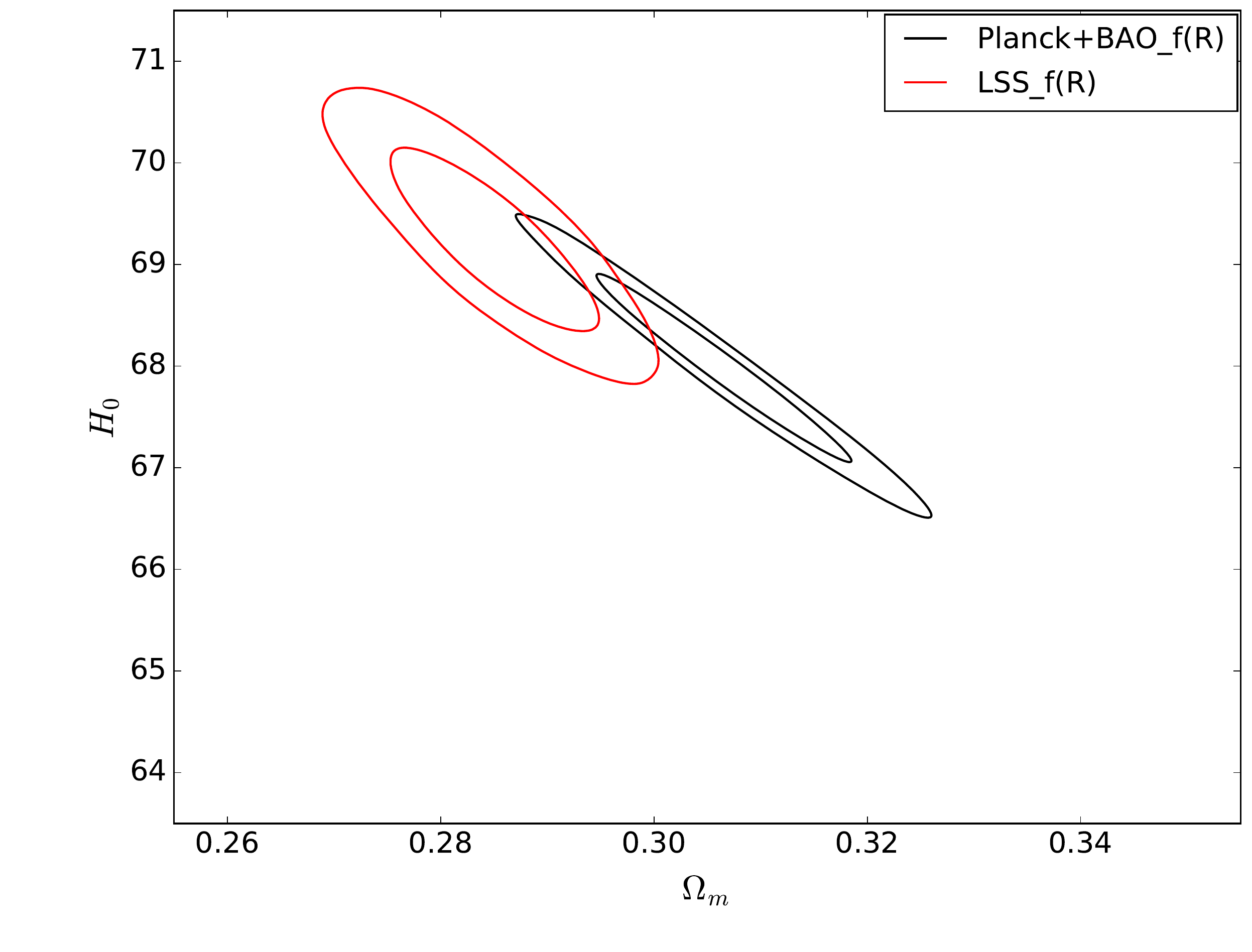}
\hspace{-0.32cm}
\includegraphics[width=2.95in,height=2.78in,angle=0]{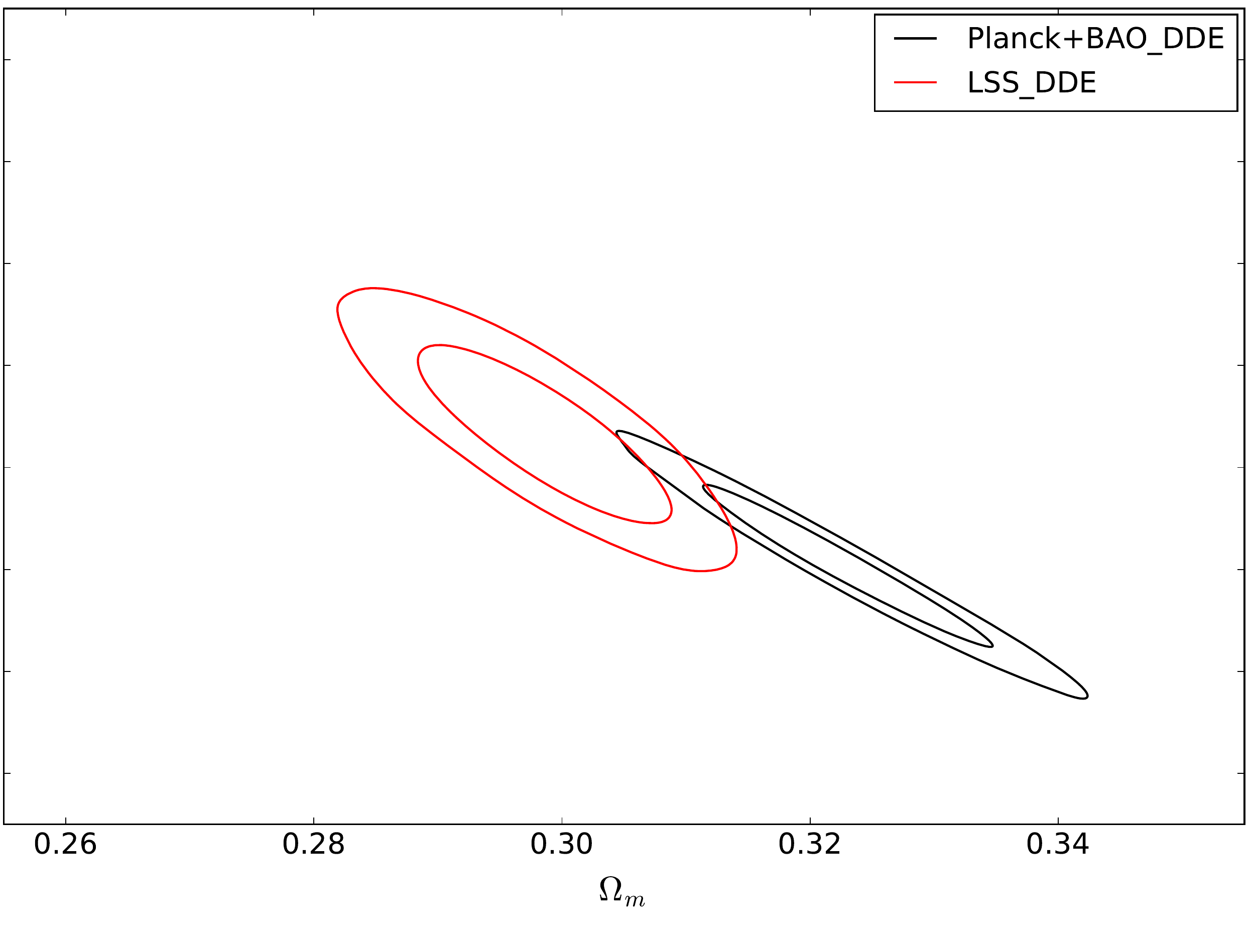} \\
(a)\hspace{6.8cm} (b) 
\caption{The 1-$\sigma$ and 2-$\sigma$ contours in $H_{0}-\Omega_m$ and $\sigma_{8}-\Omega_m$ parameter space for HS and DDE model with $\sum m_\nu=0.06$eV are shown here. Panel (a) shows that the $\sigma_8$ discrepancy worsens in the HS model whereas $H_0$ tension is slightly resolved. In panel (b) it is shown that both the tensions are slightly alleviated in DDE model. }\label{joint-nomnu}
\end{center}
\end{figure}

\section{Datasets and Analysis}
\label{sec:data}
As discussed in Sect. \ref{sec:intro} there is a discrepancy in the values of $H_{0}$ and $\sigma_{8}$ reported by the large
scale surveys and Planck CMB observations. In this paper we analyse $\Lambda$CDM, HS and DDE model. For analysing these models,
we use Planck CMB observations \cite{Ade:2015xua} for temperature anisotropy power spectrum over the multipole range 
$\ell \sim 2-2500$ and Planck CMB polarization data for 
low $\ell$ only. We refer to these data sets combined as Planck data. We also use the Baryon acoustic oscillations(BAO) data from
6dF Galaxy Survey~\cite{Beutler:2011hx}, BOSS DR11~\cite{Anderson:2013zyy,Font-Ribera:2013wce} and SDSS DR7 Main Galaxy
Sample~\cite{Ross:2014qpa}. In addition we use the cluster count data from Planck SZ survey~\cite{Ade:2013lmv},
lensing data from Canada France Hawaii Telescope Lensing Survey (CFHTLens)~\cite{Kilbinger:2012qz,Heymans:2013fya} and CMB 
lensing data from Planck lensing survey~\cite{Ade:2013tyw} and South Pole Telescope (SPT)~\cite{Schaffer:2011mz,vanEngelen:2012va}.
We also use the data for Redshift space distortions (RSD) from BOSS DR11 RSD measurements~\cite{Beutler:2013yhm}. We combine Planck SZ 
data, CFHTLens data, Planck lensing data, SPT lensing data and RSD data and refer them as LSS data. 
We perform Markov Chain Monte Carlo(MCMC) analysis
for $\Lambda$CDM, HS and DDE model with both Planck and LSS data. We use CosmoMC \cite{Lewis:2002ah} to perform the MCMC analysis
for $\Lambda$CDM and DDE model and add MGCosmoMC patch \cite{hojjati,silvestri} to it for HS model. MGCosmoMC patch 
includes the $\mu(k,a)$ and $\gamma(k,a)$ parametrization discussed in Sect. \ref{sec:st}.

\begin{figure}[!tbp]
\begin{center}
 HS model \hspace{3.5cm} DDE model \\
\hspace{-1.1cm}
\includegraphics[width=3.38in,height=2.78in,angle=0]{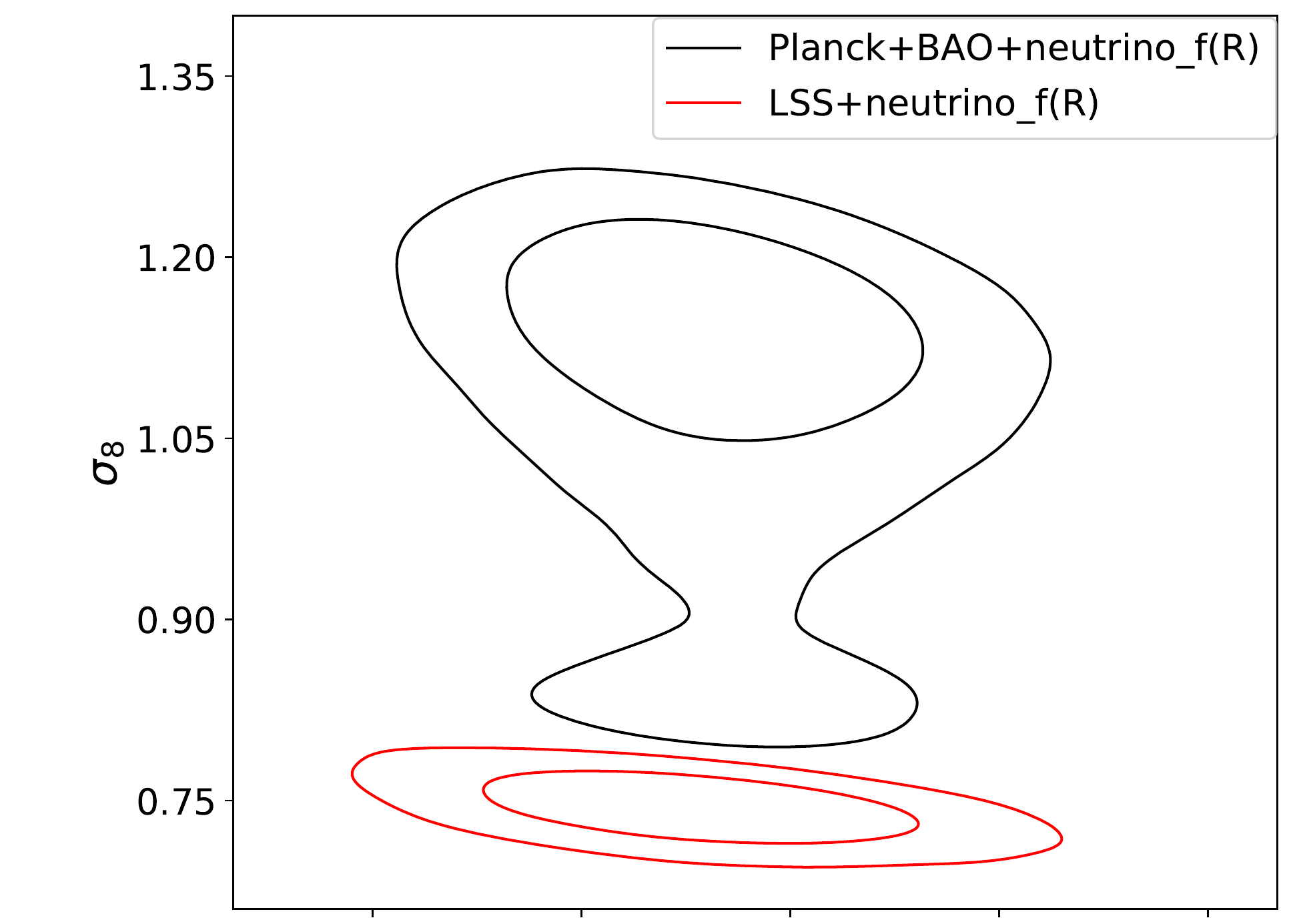}
\hspace{-0.36cm}
\includegraphics[width=2.95in,height=2.78in,angle=0]{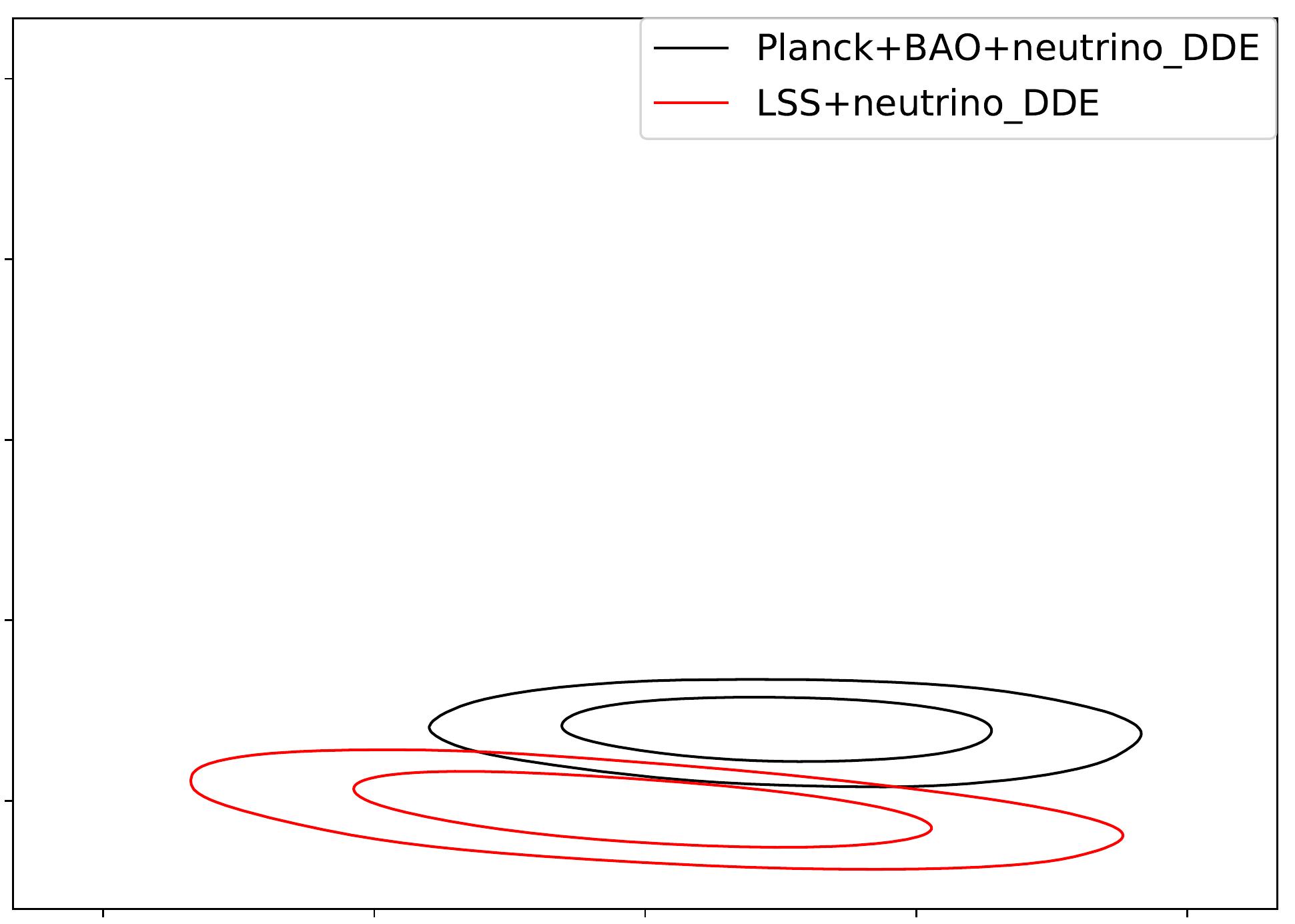}
\\
\vspace{-0.12cm}
\hspace{-1.1cm}
\includegraphics[width=3.38in,height=2.78in,angle=0]{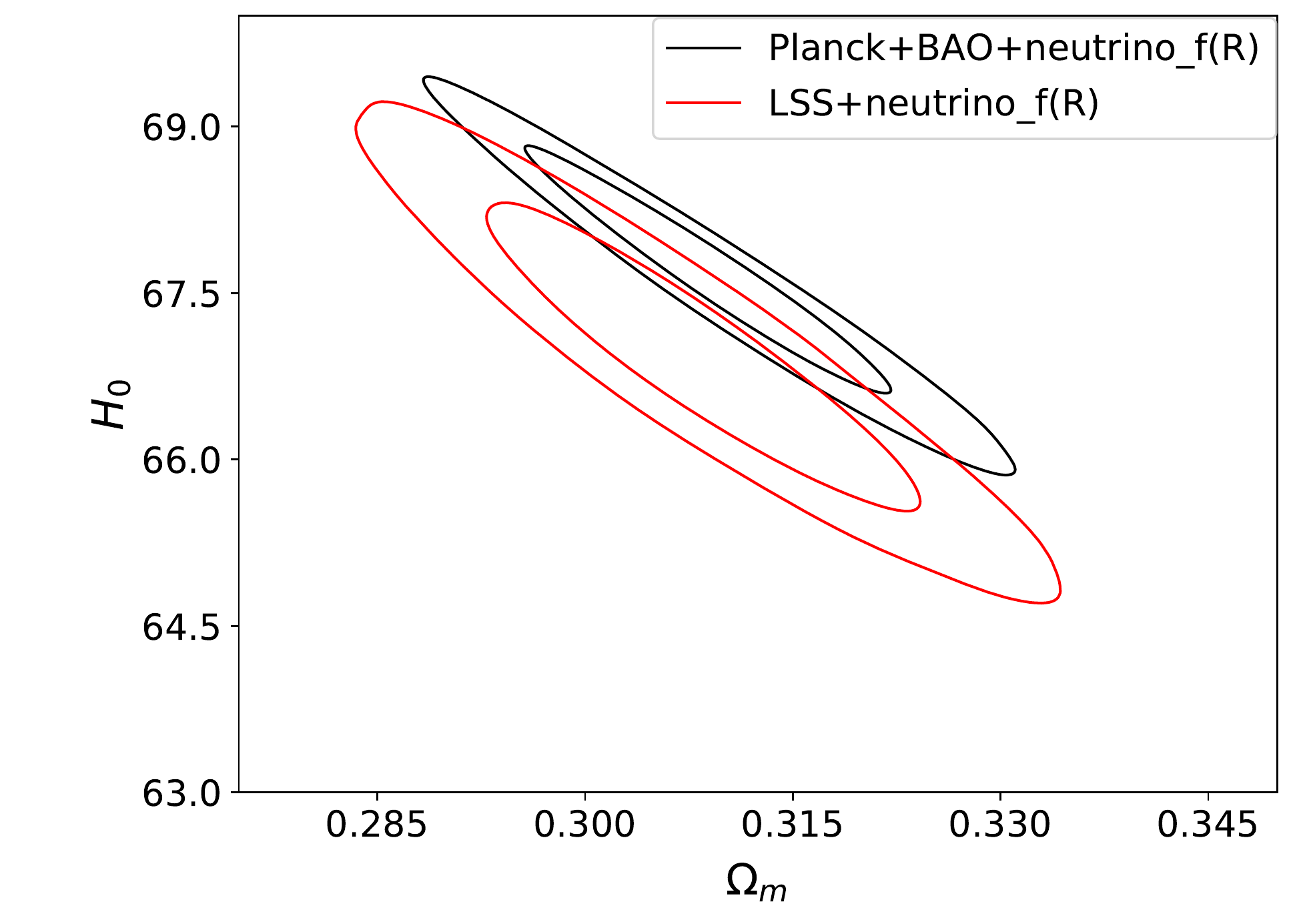}
\hspace{-0.32cm}
\includegraphics[width=2.95in,height=2.78in,angle=0]{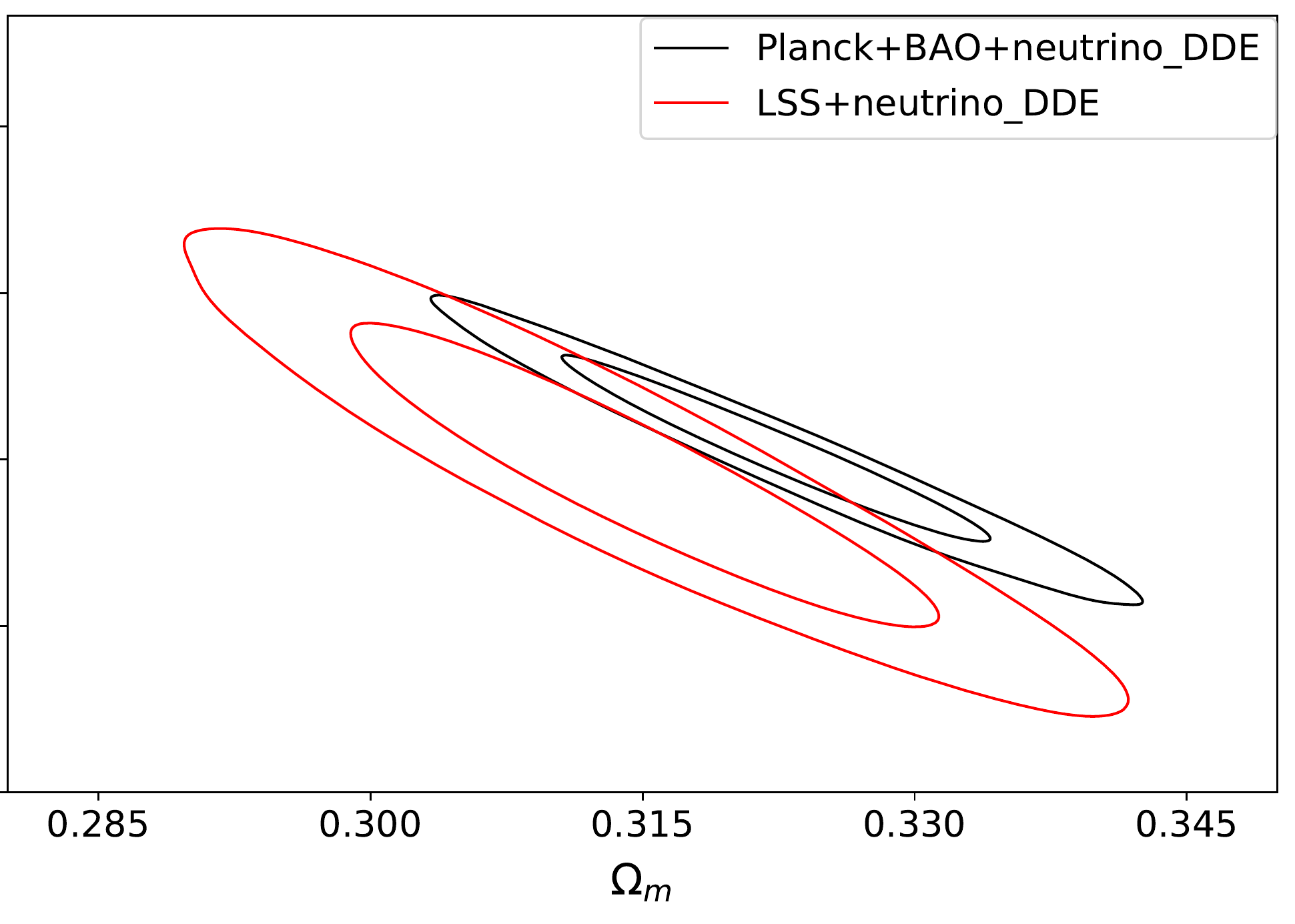} \\
(a)\hspace{6.8cm} (b) 
\caption{The 1-$\sigma$ and 2-$\sigma$ contours in $H_{0}-\Omega_m$ and $\sigma_{8}-\Omega_m$ parameter space for HS and DDE model with $\sum m_\nu$ as free parameter are shown here. Panel (a) shows that with $\sum m_\nu$ as free parameter, the $H_0$ tension is resloved and $\sigma_8$ tension is worsened. Whereas in Panel (b) it is shown that $H_0$ tension is eased and $\sigma_8$ tension is alleviated slightly. }\label{jointmnu}
\end{center}
\end{figure}

In our analysis for $\Lambda$CDM model we have total six free parameter which are standard cosmological parameters namely, 
density parameters for cold dark matter(CDM) $\Omega_{c}$ and baronic matter $\Omega_{b}$, optical depth to reionization 
$\tau_{reio}$, angular acoustic scale $\Theta_{s}$, amplitude $A_{s}$ and tilt $n_{s}$ of the primordial power spectrum. We fix 
$\sum m_{\nu} =0.06$eV to satisfy the neutrino oscillation experiments results. We also have two derived parameters $H_{0}$ and 
$\sigma_{8}$. First we perform MCMC analysis with Planck+BAO data with these parameters and get constraints for each
parameter. Next, we run the MCMC analysis with LSS data for $\Lambda$CDM model. In order to avoid the over fitting of data we 
use $\Theta_{s}=1.0413 \pm 0.0063$ and $n_{s}=0.9675\pm0.0075$, obtained from analysis with Planck+BAO data, as gaussian prior in our run of MCMC 
analysis with LSS data. Since $\tau_{reio}$ does not affects the LSS observation therefore we also use the best fit value of 
$\tau_{reio}=0.08$, obtained from analsis with Planck+BAO data, as fixed prior. These analyses
give the $H_{0}=67.7^{+0.8}_{-0.9}$ and $\sigma_{8}=0.829^{+0.021}_{-0.023}$ for the Planck+BAO data and $H_{0}=69.4^{+1.0}_{-0.9}$ and
$\sigma_{8}=0.804^{+0.009}_{-0.009}$ and for LSS data. We plot the parameter space $H_{0}-\Omega_m$ and 
$\sigma_{8}-\Omega_m$, obtained from two different analysis (fig. \ref{jointlambda}). It is clear from the 
fig. \ref{jointlambda} that there is a mismatch between the values of $H_{0}$ and $\sigma_{8}$ inferred from
Planck+BAO data and that from LSS data.
 
In our analysis for HS model we have total eight free parameter of which six are standard cosmological parameters,
two are HS model parameters namely, $f_{R_{0}}$ and $n$ as defined in Sect. \ref{sec:st}. Here we fix $n=1$ and 
allowed $f_{R_{0}}$ to vary in the range [10$^{-9}$,10]. We repeat the whole procedure to do the analysis with Planck+BAO and
LSS data for HS model and obtain constraints for each parameter. Similar to the analysis for $\Lambda$CDM
model, in the analysis of this model with LSS data, we fixed the $\tau_{reio}=0.078$ and use 
$\Theta_{s}=1.0411\pm0.00064$ and $n_{s}=0.9684\pm0.0067$ as gaussian prior for the same reason. The best fit values for $H_{0}$ and $\sigma_{8}$ 
in this analysis are $67.9^{+1.0}_{-0.9}$  and $1.097^{+0.133}_{-0.077}$ with Planck+BAO data and $69.3^{+0.8}_{-1.0}$  and 
$0.804^{+0.006}_{-0.010}$ with LSS data respectively. We plot the parameter space $H_{0}-\Omega_m$ and $\sigma_{8}-\Omega_m$, 
obtained from analysis with two different data sets, see fig. \ref{joint-nomnu}. We found that tension between the values of $\sigma_{8}$ inferred from 
Planck+BAO data and that from LSS data is increases, whereas the tension in $H_{0}$ value decreses in this model.

Similarly we do the analysis for DDE model. In our analysis for DDE model, in addition to the six standard parameters, 
we have two model parameters $w_{0}$ and $w_{a}$ as defined in Sect. \ref{sec:vde} making a total of eight parameters.
We fix $w_{0}$ and $w_{a}$ to be $-0.9$ and $-0.1$ respectively (satisfying $w_a+w_0=-1 $ to keep the field non phanton) and do MCMC 
analysis scan over the remaining six parameters. We repeat the same procedure as we did for $\Lambda$CDM and HS model.
First we do analysis with Planck+BAO data and get constraints on all the free parameters. In the analysis with LSS data, we
fix $\tau_{reio}=0.085$ and use $\Theta_{s}=1.0411\pm0.00065$ and $n_{s}=0.9601\pm0.0069$ as prior (These values are obtained in the 
analysis with Planck+BAO data). We plot the parameter space $H_{0}-\Omega_m$ and $\sigma_{8}-\Omega_m$, obtained from
analysis with two different data sets, see fig. \ref{joint-nomnu}. We find that tension between the values of $\sigma_{8}$ and $H_{0}$ values  
inferred from Planck+BAO data and that from LSS data is somewhat alleviated in the DDE model. Constraints on $\sigma_{8}$, 
$H_{0}$ and $\Omega_m$ for each model is listed in table \ref{tab:mcmc}.

\begin{table}[h!]
\hspace{-1.5cm}
\noindent\begin{tabular}{| l | c | c | c | c | c | c | c |}
\cline{3-8}
\multicolumn{1}{c}{}& \multicolumn{1}{c|}{}& \multicolumn{2}{c|}{$\sum m_{\nu}$} & \multicolumn{2}{c|}{$\sigma_{8}$} & \multicolumn{2}{c|}{$H_{0}$} \\
\cline{1-8}
& Model & {Planck+BAO } & {LSS} & {Planck+BAO } & {LSS } & {Planck+BAO } & {LSS }\\
\cline{1-8}
& $\Lambda$CDM &  &  & $0.829^{+0.021}_{-0.023}$ & $0.804^{+0.009}_{-0.009}$ &$67.7^{+0.8}_{-0.9}$ &$69.4^{+1.0}_{-0.9}$\\
\cline{5-8}
Fixed $\sum m_\nu$ &  HS & $0.06$ eV & $0.06$ eV & $1.097^{+0.133}_{-0.077}$ & $0.804^{+0.006}_{-0.010}$ & $67.9^{+1.0}_{-0.9}$ & $69.3^{+0.8}_{-1.0}$\\
\cline{5-8}
& $DDE$ &  &  &  $0.808^{+0.021}_{-0.023}$ & $0784^{+0.012}_{-0.014}$ & $66.0^{+0.8}_{-0.8}$ & $67.4^{+0.7}_{-1.0}$\\
\cline{1-8}
& $\Lambda$CDM & $\le 0.157$  & $0.379^{+0.138}_{-0.148}$ & $0.826^{+0.030}_{-0.029}$ &$0.741^{+0.031}_{-0.031}$ & $67.6^{+1.0}_{-1.0}$ & $67.2^{+1.4}_{-1.5}$\\
\cline{3-8}
Free $\sum m_\nu$ & HS& $\le 0.318$  & $0.375^{+0.131}_{-0.144}$& $1.115^{+0.114}_{-0.070}$ & $0.744^{+0.029}_{-0.030}$ & $67.6^{+1.2}_{-1.0}$ & $66.9^{+1.4}_{-1.4}$\\
\cline{3-8}
& $DDE$ & $\le 0.116$  & $0.276^{+0.139}_{-0.155}$ & $0.809^{+0.026}_{-0.027}$ & $0.742^{+0.032}_{-0.031}$ & $66.0^{+0.9}_{-0.8}$ & $65.9^{+1.3}_{-1.4}$\\
\cline{1-8}
\end{tabular}

\caption{The best fit values with 1-$\sigma$ error for $\sum m_\nu$, $\sigma_8$ and $H_0$ obtained from the MCMC analyses for all the
models considered are listed here.}\label{tab:mcmc}
\end{table}

\begin{figure}[t!]
\begin{center}
Planck+BAO \hspace{6.5cm} LSS \\
\hspace{-1.1cm}
\subfloat[\label{fig:frmnu}]{
\includegraphics[width=2.95in,height=2.8in,angle=0]{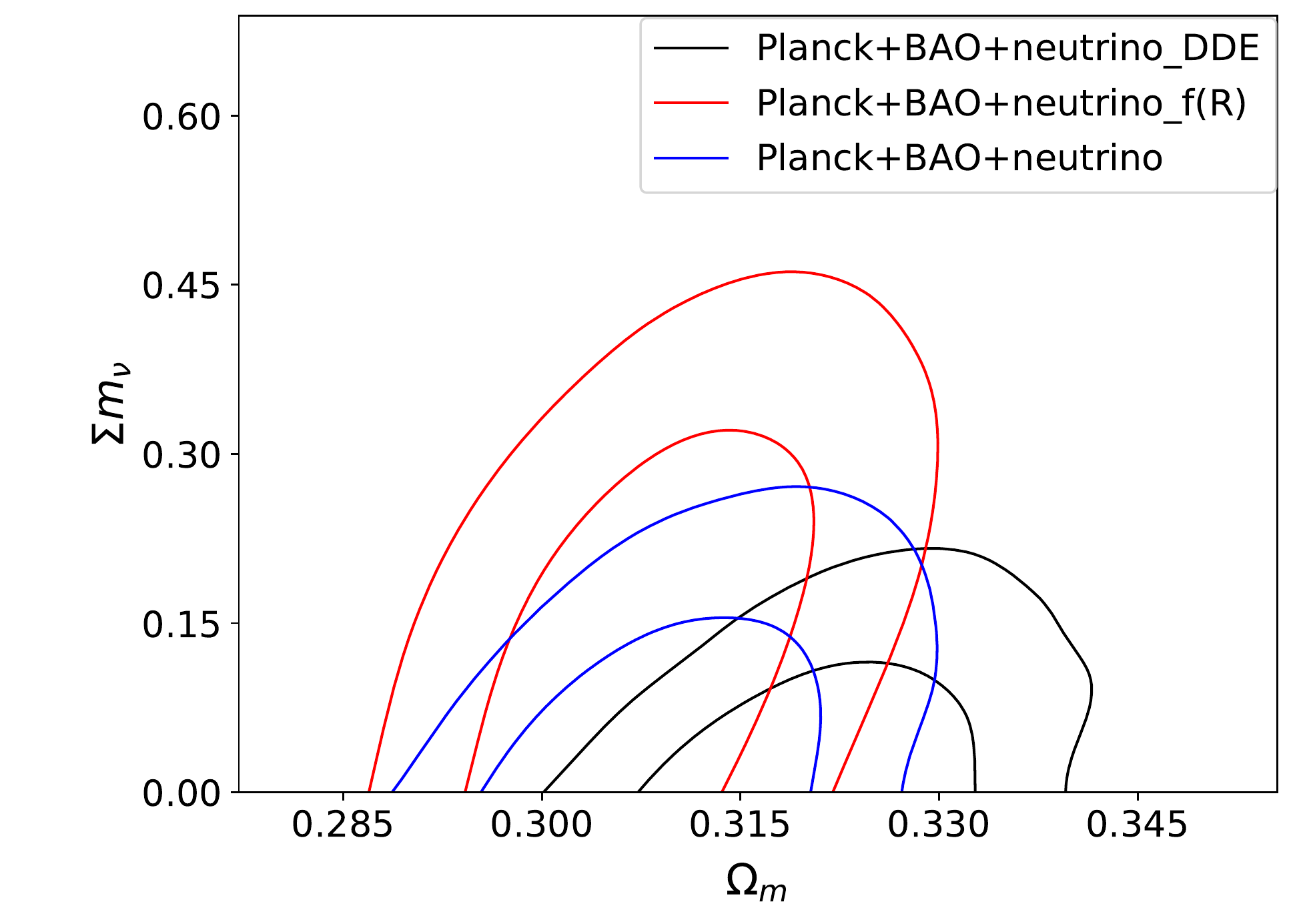}}
\hspace{-0.32cm}
\subfloat[\label{fig:vdemnu}]{
\includegraphics[width=3.38in,height=2.78in,angle=0]{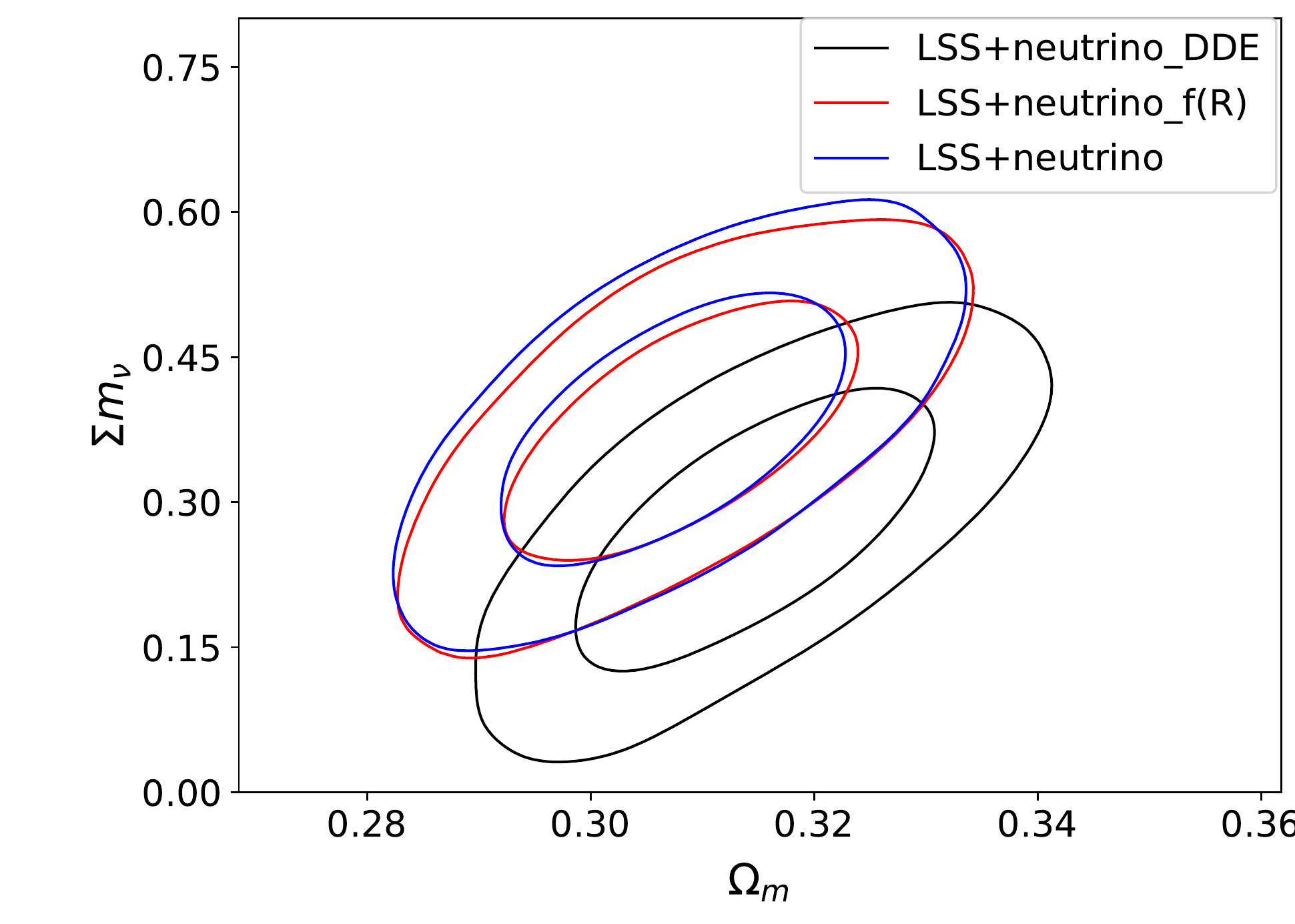}}
\caption{Bounds on the Neutrino mass in DDE, HS and $\Lambda$CDM models with Planck and LSS data.}\label{nubound}
\end{center}
\end{figure}

Next, we use sum of massive neutrino $\sum m_{\nu}$ as a free parameter and allow it to vary in the range [0,5]eV in our
analysis for all three models. 
We repeat the whole procedure and obtain constraints for each parameter. We  plot the parameter space $H_{0}-\Omega_m$ and
$\sigma_{8}-\Omega_m$ for each model, see fig. \ref{jointlambda} and \ref{jointmnu}. The constraint on $\sum M_{\nu}$ in $\Lambda$CDM, HS and DDE model is listed in table \ref{tab:mcmc}. 
The corresponding $1\sigma$ and $2\sigma$ contours are shown in fig. \ref{fig:frmnu} and \ref{fig:vdemnu}. 
We also list the constraints on $\sigma_{8}$ and $H_{0}$ in each model with Planck+BAO and LSS data, in table \ref{tab:mcmc}.

\section{Discussion and Conclusion}\label{sec:conc}
Galaxy surveys and CMB lensing measure the parameter $\sigma_8 \Omega_{m}^{\alpha}$, where $\Omega_{m}^{\alpha}$ represents a model
 dependent growth function. In $\Lambda$CDM $\alpha=0.3$ but it could be different for other DM-DE models. In CMB measurement of temperature anisotropy spectrum $C_l$ and BAO determine $\Omega_m$. 
The discrepancy between the CMB and LSS measurement is determined by the model dependent growth function $\Omega_{m}^{\alpha}$.
The growth function can thus be used for testing theories of gravity and dynamical DE. In the present paper we tested HS and DDE models in the context of $\sigma_8-\Omega_m$ observations. We find that in the HS model the $\sigma_8-\Omega_m$ tension worsens compared to the $\Lambda$CDM model. On the other hand in the DDE model there is slight improvement in the concordance between the two data sets. The bounds on neutrino mass become more stringent in the DDE model. In the HS model there is a loosening in the analysis with Planck data and not much effect in the analysis with the LSS data.

In conclusion we see that $\sigma_8$ measurement from CMB and LSS experiments can be used as a probe of modified gravity or quintessence models. Future observations of CMB and LSS may shrink the parameter space for $\sigma_8-\Omega_m$ and then help in selecting the correct $f(R)$ and DDE theory.

\bibliographystyle{JHEP}
\bibliography{LSSnuMASS.bib}  
  
\end{document}